\documentclass[useAMS,usenatbib]{mn2e}
\usepackage{epsfig,amssymb,amsmath,graphicx}

\title[Broad Band X-ray Spectra of 4U 1636-536]{Broad Band X-ray Spectra of Atoll Source 4U 1636-536: NuSTAR and Swift Results}
\author[V. K. Agrawal and Mohammad Hasan]{V. K. Agrawal$^{1}$\thanks{E-mail:
vivekag@isac.gov.in} and Mohammad Hasan$^1$ \\
$^{1}$ Space Astronomy Group, ISRO Satellite Center, ISITE Campus,Bangalore 560037,INDIA \\
}
\begin{document}


\pagerange{\pageref{firstpage}--\pageref{lastpage}} \pubyear{2002}

\maketitle

\label{firstpage}

\begin{abstract}

In this work we investigate  broad band (1-79 keV) spectral nature of the
atoll source 4U 1636-536 using the combined  Nu-STAR and SWIFT-XRT data. The
spectra are complex and have emission components from the disc, boundary-layer
and corona. In addition to that a broad iron line is also observed. A
relativistic line model assuming Schwarzchild metric fits this
feature. The total flux varies from 1.4 $\times$ 10$^{-9}$ to  4.36
$\times$ 10$^{-9}$ $ergs/s/cm^2$. At the highest flux level the source
was found  in the soft state. In this state the Comptonized component
has temperature $kT_e \sim 3$ keV and optical depth  $\tau \sim $ 16. We
also detect a non-thermal tail with index $\sim$ 2.4, contributing $\sim$
10 \% of the total flux in the soft state.  We also find that the inner disc
radius, electron temperature and optical depth vary with the total
0.1-100 keV unabsorbed flux. We discuss the implication of the results
in this paper.
\end{abstract}

\begin{keywords}
accretion, accretion discs - X-rays: binaries - X-rays: individual: 4U 1636-536
\end{keywords}

\section{Introduction}
Low magnetic field neutron stars in low-mass X-ray binaries (LMXBs) are of great
interest due to the fact that they provide an excellent laboratory to study
the properties of matter at extreme densities. They are broadly divided
in two classes: Atoll source and Z-source based on their correlated spectral
and timing properties \citep{Hasvan89}.

Atoll sources trace fragmented path in the Color-Color diagram (CD).
They have luminosity in the range of 0.01-0.2 $L_{Edd}$ \citep{Done07}. The atoll-track consists of an upwardly curved branch at the right hand side of the CD, called  `banana' branch and bunch of uncorrelated points in the left hand side called `island state'.

The source 4U 1636-536 is one of the well studied atoll source.  The source has exhibited Quasi-Periodic-Oscillations \citep{Wij97}.  type-I X-ray bursts \citep{Galloway06} and millisecond oscillations during  thermonuclear bursts \citep{Strohmayer02}.  The orbital period of the source is $\sim$ 3.8 hr \citep{Giles02}. The optical observations suggest that inclination lies in the range $\sim 30-60^{\circ}$ \citep{Casares06}. The distance to the source has been estimated to be 6.0$\pm$0.5 kpc \citep{Galloway06}.

X-ray spectra of  neutron star LMXBs are in general complex and
have multiple emission components, arising in different regions of
the accretion flow. The complex spectra have two main components, soft
and hard. Two different approach have been adopted to fit the X-ray
spectra of neutron star LMXBs. In first approach, the soft component  is
modeled  as multicolor-disc-blackbody (MCD) \citep{Mit84,Mit89}, coming
from optically thick accretion disc and the hard component is described by
Comptonized emission from  the boundary-layer \citep{agrawal03,agrawal09,
disalvo02,Barret01,Barret00,Tarana08}.  In second approach, the soft
component is modeled as blackbody emission from the hot surface of neutron
star and the hard Comptonized component is considered to be arising from
the hot inner flow \citep{disalvo00,Barret01,Barret00,Sleator16}. However,
combination of two thermal components (Blackbody and MCD) and Comptonized
emission have also been used to model the X-ray spectra of atoll sources
Ser X-1 \citep{Chiang16,Miller13} and 4U 1636-536 \citep{Sanna13,Lyu14}.

Spectra of Z-sources are generally soft. The Comptonized component has
temperatures in the range of 2-5 keV and the optical depth in the range of
10-20 for a spherical geometry of the corona \citep{agrawal09,Barret01}.
Atoll sources exhibit two types of state: hard and soft. In the soft
state, spectra are  similar to the Z-sources. However, in the hard state,
the temperature of Comptonized component has a value in the range of
10-50 keV and some time extend beyond 100 keV \citep{Piraino99}. In the hard state optical depth of the corona is found to be in the range of $\sim$ 2-4 for a
spherical geometry of the corona \citep{Barret01,Barret00,Tarana08}.
During the soft state of  atoll sources a  hard power-law tail has
also been observed \citep{Pirano07,Tarana11}.  

A part of  hard X-rays
from corona illuminates the accretion disc and gets reprocessed and
reflected. The reflected emission consists of a Compton hump that
peaks  $\sim$ 10-40 keV and a relativistically smeared iron K$_\alpha$
line \citep{Fabian89, Miller07,Reynolds03}. Relativistically smeared
iron line have  been observed in a few NS LMXBs with NuSTAR (eg. 4U
1728-34,\citealt{Sleator16}; Serpens X-1 \citealt{Miller13}; 4U 1608-522
\citealt{Degenaar15}).

In this paper we present broad band spectra of neutron star
low-mass X-ray binary 4U 1636-536. The broad spectra are analyzed
using quasi-simultaneous data from SWIFT-XRT (1-10 keV) and NuSTAR
(3-79 keV). In section 2 we  present observation and data reduction
procedure. In section 3 we provide analysis methodology and highlight
the results. In section 4  we discuss the results of the spectral analysis.

\section{Observation and Data Reduction}
\begin{table}
\caption{NuSTAR and Swift observations of 4U 1636-536}
\begin{tabular}{ccccc}
\hline
\hline
Obs &  ObsID          & Start Date   & START UT& Duration \\
& (NuSTAR) &(hh:mm:ss)& (ksec) \\
\hline
1 & 30101024002  &2015-06-10& 08:11:07& 19.8\\
2 & 30102014002  &2015-08-25  &  02:51:08 & 27.4\\
3 & 30102014004  &2015-09-05  &  17:41:05 & 27.4\\
4 & 30102014006  &2015-09-18  &  07:06:10 & 27.4\\
\hline
Obs & ObsID             & Start Date   & START UT& Duration \\
&(SWIFT)&(hh:mm:ss)& (ksec) \\
\hline
1 & 00081640002 & 2015-06-10 & 10:10:43  &  1.47\\
2 & 00081640003 &2015-08-25 & 14:02:52   & 2.75\\
3 & 00081594001 & 2015-09-05&21:15:52 &0.7 \\
4 & 00081594002 & 2015-09-18 & 14:16:31 & 2.9 \\
\hline
\hline
\end{tabular}
\end{table}

{\it NuSTAR} observed the atoll source 4U 1636-536  four times during
June 2015 to Sep 2015. The observation log is shown in Table 1. The
source was observed for a total duration of  $\sim$ 106 ks. The data
were processed using the NuSTAR Data Analysis Software (NuSTARDAS).
The clean event files were created  using {\it nupipeline}. A circular
region of size 80$\arcsec$  centered at the source position was created
to extract the source spectra and light curves.  The background events
were extracted from a circular region away from the source with size
similar to the source region. The standard FTOOL task {\it nuproducts}
was used to create the spectra, lightcurves and response files.

Quasi-simultaneous Swift XRT data were also available in the archive (see
Table 1). The swift data were taken in ``windowed timing mode". We produced
clean event and area response files using XRTPIPELINE software. From the
clean events source events were extracted from a circular region of 20
pixels. The background events were extracted for an annular region with
inner radius of 20 pixels and outer radius of 180 pixels. The spectra
and light curves were extracted using XSELECT software. The backscale
was set to 40  (twice the radius of circular source region) for the
source spectra and to 160 (difference between inner and outer radius of
annular region) for the background spectra. The count rate was below 100
counts/s for all the observations and hence spectra were pileup free.
\begin{figure*}
\centering
\begin{tabular}{@{}cc@{}}
\hspace{-.4in}
\includegraphics[width=0.35\textwidth,angle=-90]{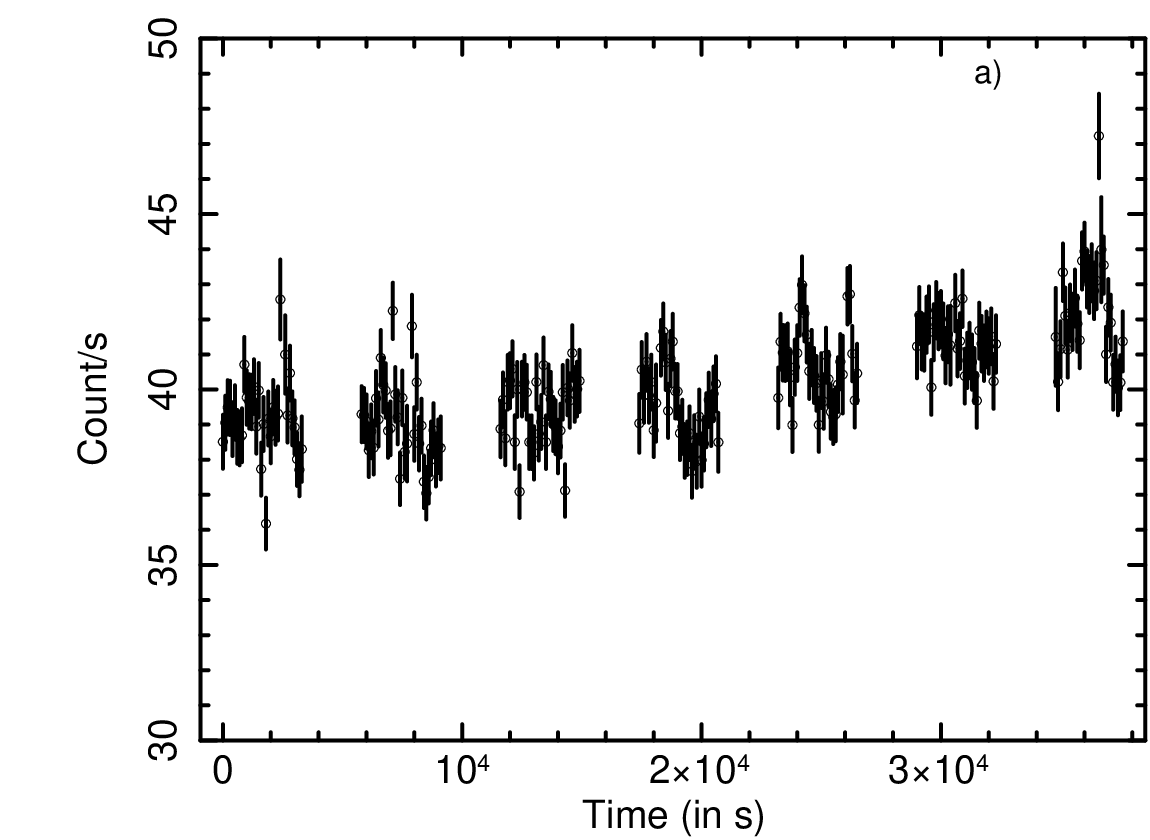}&
\includegraphics[width=0.35\textwidth, angle=-90]{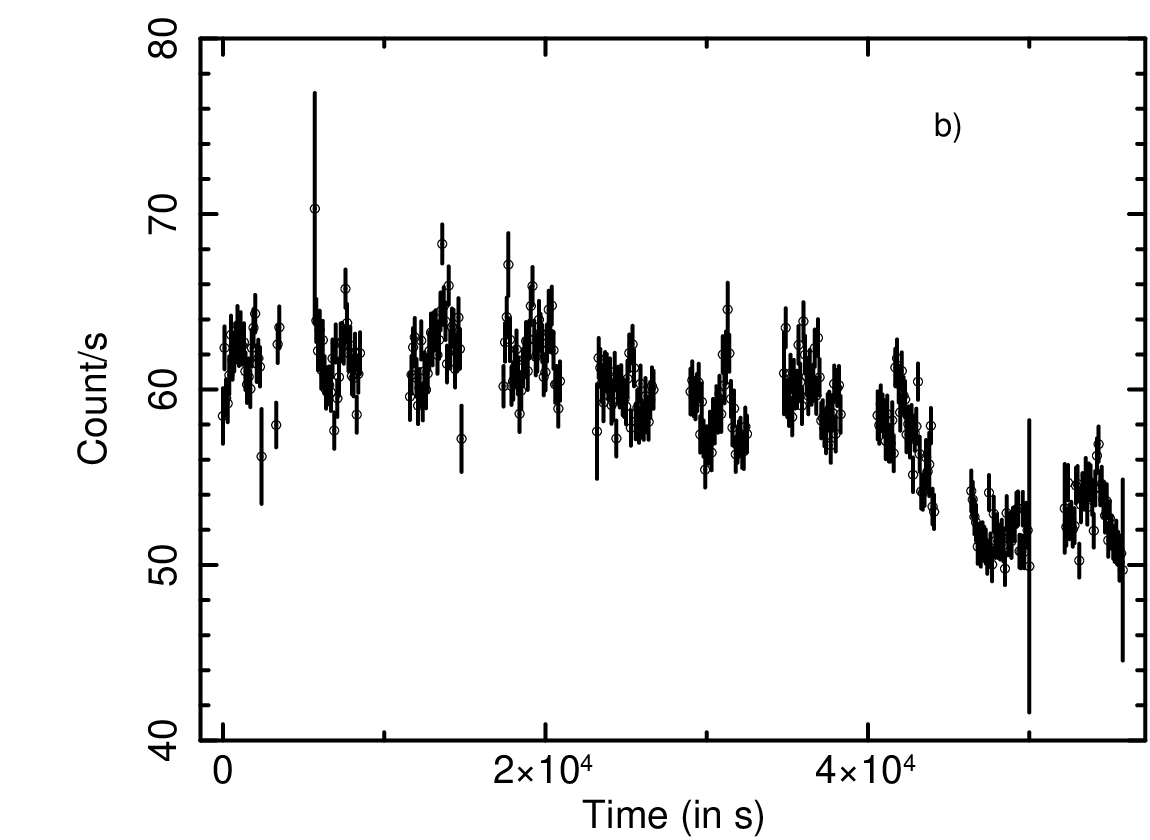} \\
\hspace{-.4in}
\includegraphics[width=0.35\textwidth,angle=-90]{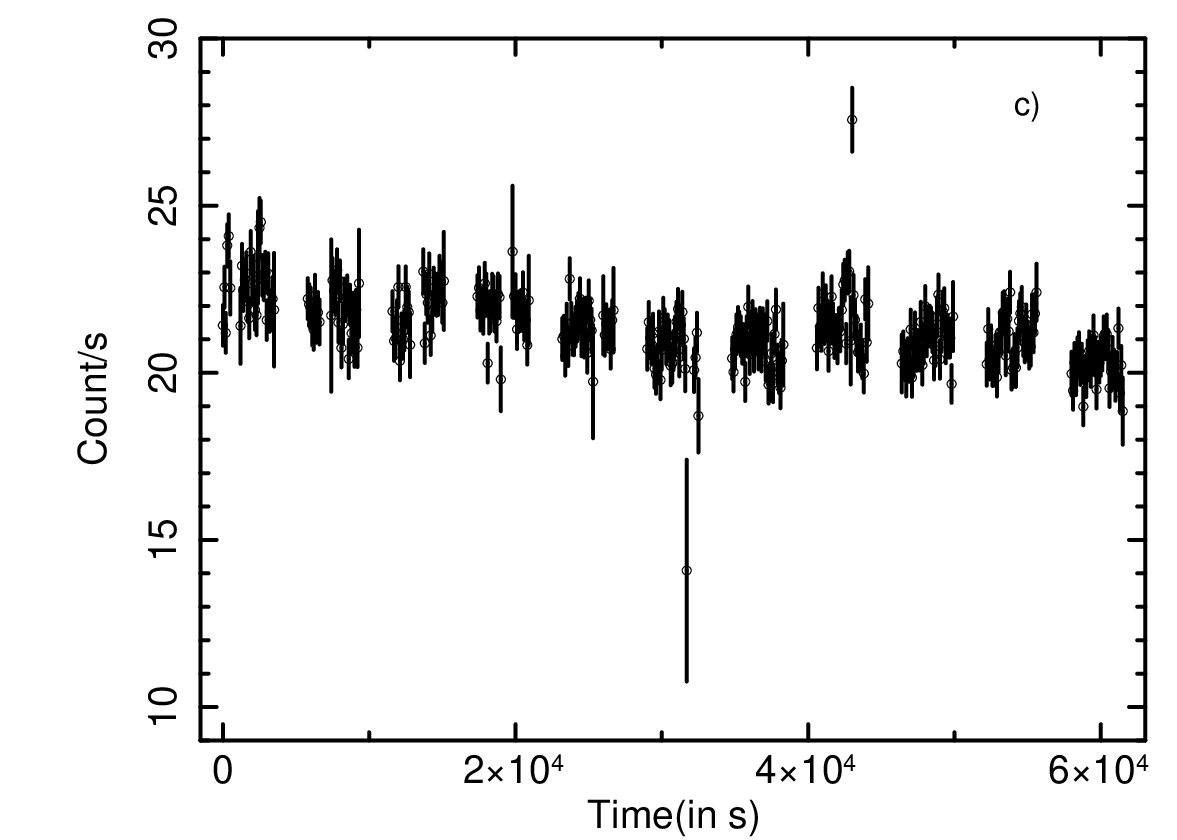}&
\includegraphics[width=0.35\textwidth,angle=-90]{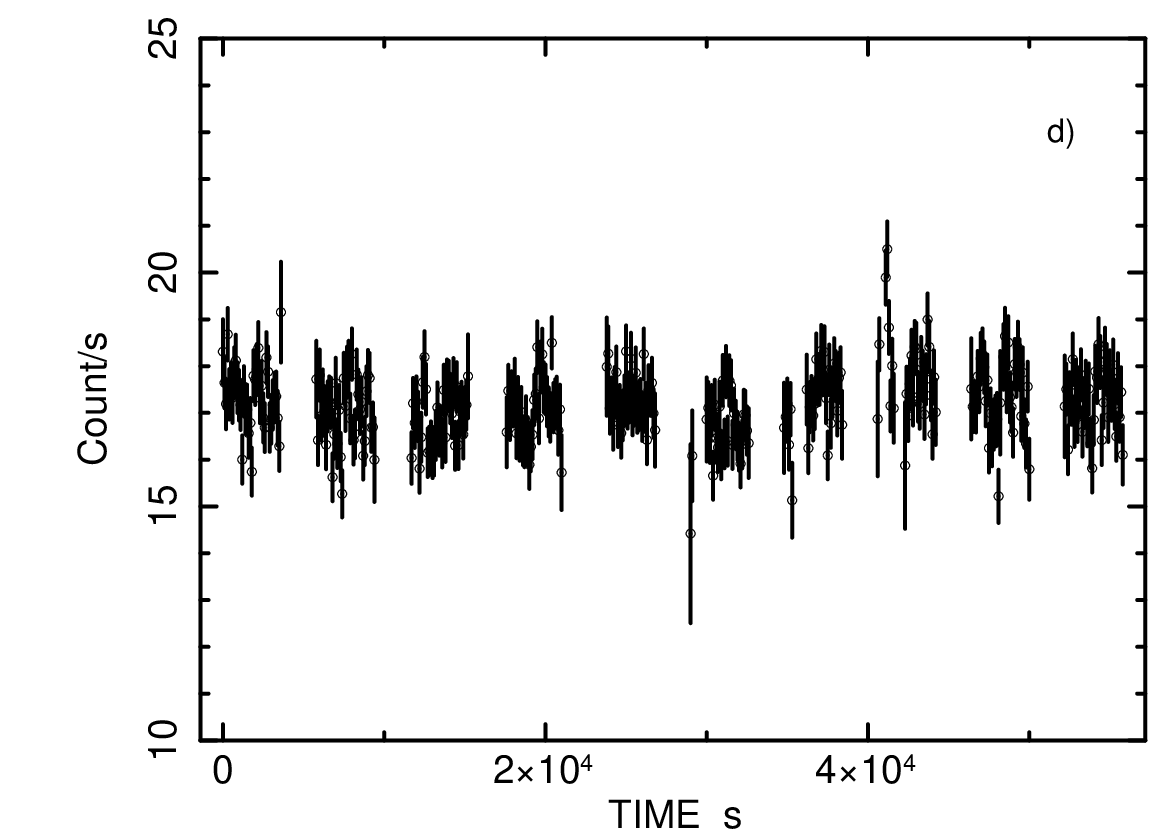} \\
\end{tabular}
\caption{Lightcurves of the atoll source 4U 1636-536 as observed by NuSTAR. Figures a),b),c) and d)  are the lightcurves for Obs-1,Obs-2,Obs-3 and Obs-4 respectively . The time bin used is 100 seconds. It is clear that the source is in the high-intensity state during Obs-2 (see Figure b above) }
\end{figure*}

Simultaneous XRT and NuSTAR spectra were fitted with spectral analysis package XSPEC \citep{Arnaud00}. We used the latest calibration files  to generate the response matrices. The spectra were grouped to give a minimum of 50 counts/bin. Unless quoted explicitly the error bars on the spectral parameters are computed using $\Delta \chi^2$ = 1.0.

\begin{table*}
\caption{In this table we give our approch to obtain the correct spectral model which fits the combined NuSTAR and Swift spectra.}
\begin{tabular}{ccccc}
\hline
\hline
Model (tried)  & $\chi^2_\nu$(Obs-1)     &  $\chi^2_\nu$(Obs-2)  & $\chi^2_\nu$(Obs-3) & $\chi^2_\nu$(Obs-4) \\
\hline
diskbb+nthComp & 4.02                    & 1.58 & 2.21  & 1.24\\
diskbb+nthComp + bbodyrad & 1.16 & 1.18 &   1.25 & 1.07 \\
diskbb+nthComp + bbodyrad+gauss & 1.02    &1.12     &1.14  & 1.02\\
diskbb+nthComp + bbodyrad+diskline & 1.05 &1.14     &1.16  & 1.02\\
diskbb+nthComp + bbodyrad+Gauss+power-law & - &1.07 &- & -\\
diskbb+nthComp + bbodyrad+diskline+power-law &-& 1.08 & -&-\\
\hline
\end{tabular}
\end{table*}
\begin{table*}
\caption{In this table we give  best-fit spectral fit parameters for
combined NuSTAR-SWIFT observations. $kT_{in}$ is inner disk temperature
and $N_{dbb}$ is normalization in XSPEC model $diskbb$. $R_{disk}$
is inner disc radius obtained using $diskline$ model. $kT_{BB}$
is blackbody temperature and $N_{BB}$ is normalization of $bbodyrad$ 
model. The parameters $kT_e$ is electron temperature and $\Gamma_c$
is power-law  photon-index  obtained from $nthComp$ model. $\tau$ is
optical depth estimated using $\Gamma_c$ and $kT_e$. $FWHM$ is full
width-half maxima of Gaussian component. $EW$ is equivalent width of
diskline component. $F_{tot}$  , $F_{Compt}$,
,$F_{pow}$ , $F_{dbb}$ $F_{bb}$ are total,Comptonized,power-law,disk-blackbody and blackbody 0.1-100 keV unabsorbed fluxes. All the fluxes  are in the units of $10^{-9} ~ ergs/s/cm^2$. }
\begin{tabular}{ccccc}
\hline
\hline
Parameters & Obs-1 & Obs-2 & Obs-3 & Obs-4 \\
$kT_{in}u$ (in keV) & 0.56$\pm$0.01 & 0.87$\pm$0.02 & 0.52$\pm$0.02 &0.73$\pm$0.04 \\
$N_{dbb}$  & 388$\pm$18.6 & 204$\pm$7.5 & 320$\pm$23.2 & 49$\pm$2.1 \\
$R_{disk}$(in $R_g$) &   8.89$\pm$1.19 & 6.8$\pm$0.5 &14.95$^{+8.9}_{-2.7}$ &24.9$\pm$4.6\\
$kT_{BB}$ (in keV)    & 1.69$\pm$0.02 &1.54$\pm$0.03 &1.39$\pm$0.02 &1.23$\pm$0.04 \\ 
$N_{BB}$   &3.4$\pm$00.41 &15.45$\pm$0.42 &4.6$\pm$0.45 &2.66$\pm$0.35 \\ 
$\Gamma_c$   &2.03$\pm$0.02&1.47$\pm$0.03 &1.94$\pm$0.03 &1.76$\pm$0.01 \\ 
$kT_e$  &10.1$\pm$0.39 &2.93$\pm$0.03& 13.16$\pm$0.71 & 19.05$\pm$1.24\\ 
$\tau$ & 4.65$\pm$0.11 &16.45$\pm$0.72 &4.30$\pm$0.18  &4.03$\pm$0.29 \\
$FWHM$ (in keV) & 1.1$\pm$0.05 & 1.44$\pm$0.04&1.39$\pm$0.12&1.14$\pm$0.16\\
 EW(in eV) & 160 & 120 & 108 &102 \\
$F_{tot}$    &3.09$\pm$0.12&4.36$\pm$0.11 &1.95$\pm$0.05&1.42$\pm$0.04 \\
$F_{Compt}$    &1.81$\pm$0.09&0.67$\pm$0.08&1.04$\pm$0.04&0.977$\pm$0.08 \\
$F_{dbb}$    &0.776$\pm$0.03&2.04$\pm$0. 05&0.457$\pm$0.03&0.213$\pm$0.01 \\
$F_{bb}$    &0.436$\pm$0.02 & 1.20$\pm$0.06&0.438$\pm$0.02&0.239$\pm$0.02 \\
$F_{pow}$    & -&0.416$\pm$0.009 &- & -\\
\hline
\end{tabular}
\end{table*}

\section{Analysis and Results}

The investigation of background subtracted lightcurves revealed the presence of
13 type-I X-ray bursts. We have shown the lightcurves of all the observations after removing the type-I X-ray bursts for a comparison of persistent flux level (see Figure 1). We exclude the type-I X-ray bursts and fit the
continuum spectral model to the persistent emission. We use the $tbabs$
model \citep{Wilms00}  to account for the interstellar absorption in the
direction of  4U 1636-536. The neutral hydrogen column density (nH) in the direction of 4U 1636-536 is $\sim$ 4.1 $\times$ 10$^{21}$ and hence it is fixed at this value. The spectral fit is performed for the combined
Swift-XRT data (1-9 keV) and NuSTAR spectra (3-79 keV).  The Swift-XRT is
poorly calibrated between  the energy range 1.6-2.1 keV, hence we ignore
the data between this energy range while performing the spectral modeling.

First we fit the combined data using the $diskbb$ and the thermal
Comptonization model. Basic thermal Comptonization models in
XSPEC are : $CompST$ \citep{Titarchuk80}, $CompTT$ \citep{Titar94} and
$nthComp$ \citep{Zdz96}. The seed photons for inverse Compton
scattering can be either single temperature blackbody emission arising
in the surface of Neutron-Star/Boundary-Layer or Multi-color-disc component
(MCD) from standard accretion disc. We use the $nthComp$ model in XSPEC
\citep{Zdz96,Zycki99} to describe the Comptonized
emission from the corona. The reason for choosing $nthComp$ model is that,
this emission model provides a way to choose the source of the seed
photons.  We choose MCD as source of seed photons and tie the value of
the seed photon temperature with the inner disc temperature. Then we add 
a single temperature blackbody ({\it bbodyrad} in XSPEC)  component to account for the emission from the neutron star
surface. The addition of this component improves the fit considerably.
There is still some residual left around 4-7 keV suggesting Fe line
component. An addition of Gaussian line component $\sim$ 6.4-6.7 keV improves
the fit. We adopted similar procedure for other observations. We also
try $diskline$ model (see Fabian et al 1989) for a Schwarzschild metric
to describe the broad iron line feature. This procedure was repeated
for all the observations and results are given in Table 2.
In Figure 2 we show the unfolded spectra for observation 1.
  For the
second observation (OBSID 30102014002), a significant residual above 30 keV is observed (see Figure 3a ).
This suggests the presence of an additional hard component in the spectrum. Hence we add a
power-law component to remove this residual (see Figure 3b ). This further improves the fit.
Therefore we consider $diskbb+nthComp+bbodyrad+diskline$ model as final description of the combined NuSTAR and SWIFT-XRT spectra except for OBSID 30102014002 where an additional power-law component is required. 

\begin{figure*}
\includegraphics[width=0.55\textwidth,angle=-90]{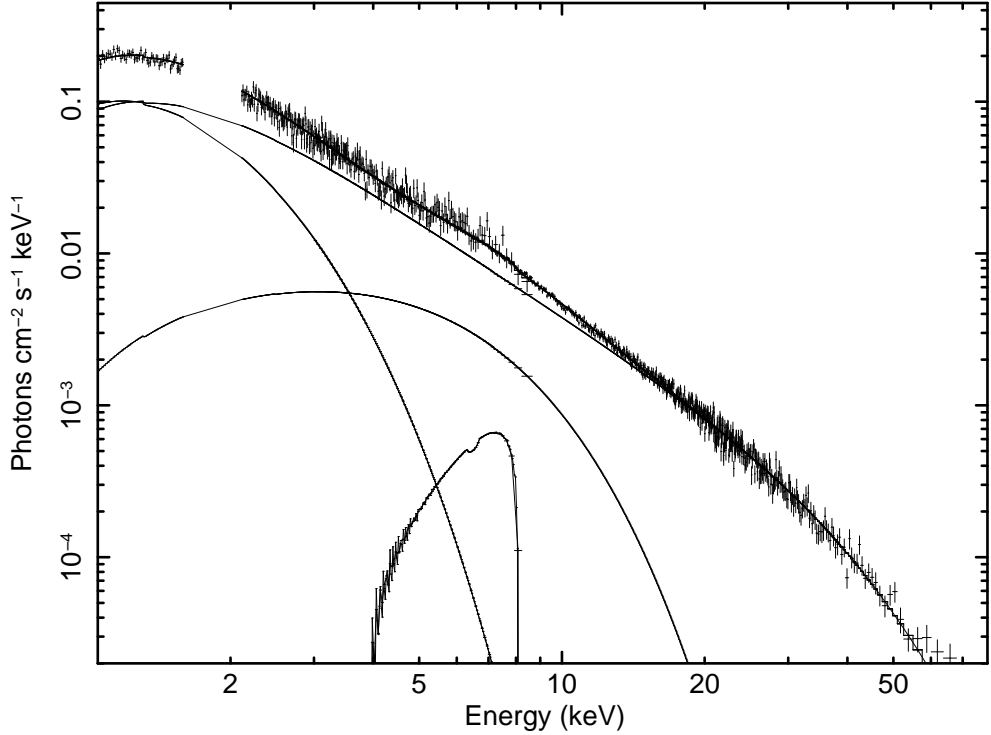}
\caption{The unfolded spectrum for Obs-1. The spectral components are: {\it diskbb},{\it diskline} {\it bbodyrad} and {\it nthComp}.}
\end{figure*}
\begin{figure*}
\centering
\begin{tabular}{@{}cc@{}}
\hspace{-.4in}
\includegraphics[width=0.32\textwidth,angle=-90]{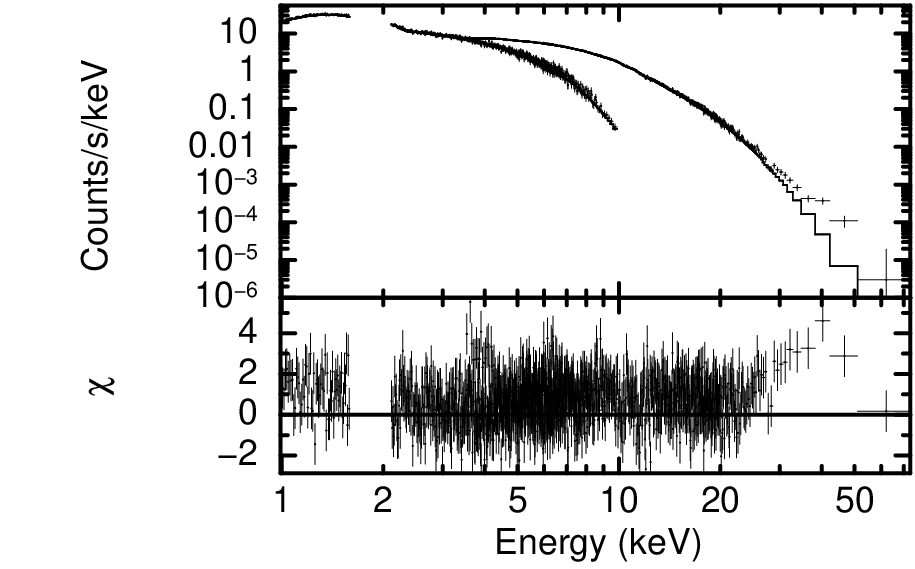}&
\includegraphics[width=0.32\textwidth, angle=-90]{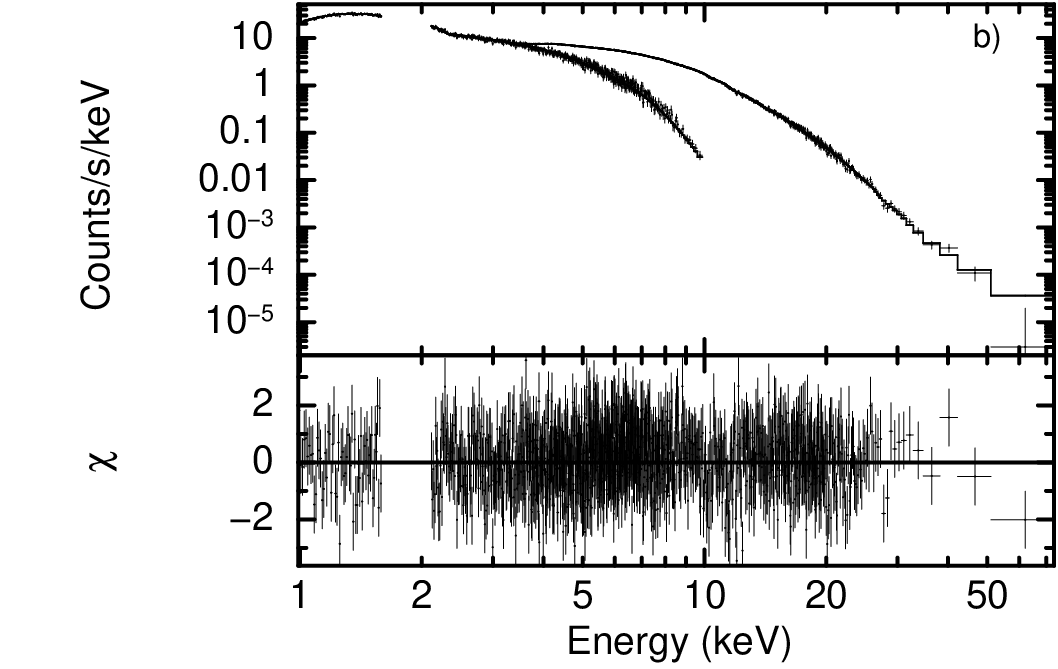} \\
\hspace{-.4in}
\end{tabular}
\caption{a) The data and best fit folded model for obs-2. It is clear from this figure that there is residual above 25 keV. b) Inclusion of power-law components remove the residual and improves the fit.}
\end{figure*}
\begin{figure}
\includegraphics[width=0.32\textwidth,angle=-90]{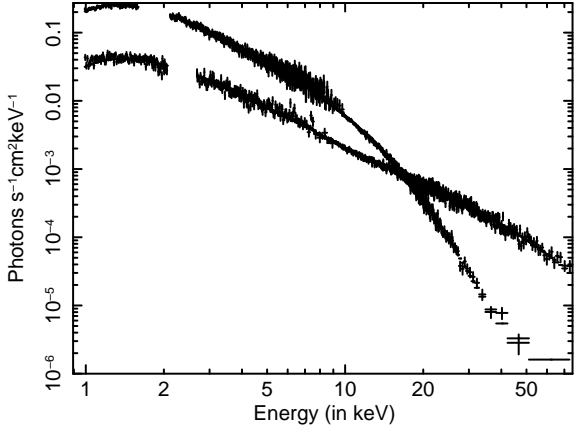}
\caption{ In this figure we plot spectra during soft (Obs-2) and hard (Obs-4) state for the comparison}
\end{figure}
\begin{figure}
\includegraphics[width=0.32\textwidth,angle=-90]{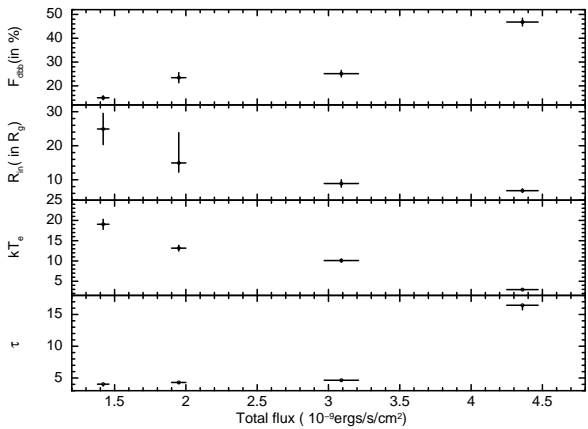}
\caption{Variation of the best fit spectral parameters as function of total (0.1-100 keV) unabsorbed flux. In panel 1 we have also plotted the percentage contribution of disc to the total flux in 0.1-100 keV band.}
\end{figure}

The best fit spectral parameters for the final model are listed in
Table 3.  The source exhibits both soft and hard states during NuSTAR
and SWIFT observations. During  Obs-2 the source is found to be  in the
soft state. For this observation the  corona  temperature $kT_e$ is $ \sim 3$
keV and the optical depth $\tau$ is $ \sim 16$ . An additional non-thermal
tail with photon index $\Gamma \sim 2.45$ is also present in this
observation. For other three observations $kT_e$ is $\sim 10-20$ keV
and $\tau$ is $\sim 4-5$.

The temperature of the soft disc component is found to be
in the range of $\sim$ 0.5-0.9 keV. During the soft state the disc
temperature is the highest (0.87$\pm$0.02 keV) suggesting that the
disc has moved closer to the neutron star surface. This is also evident
from the fact that the inner disc radius derived from the $diskline$ is
$\sim$ 6.8 $GM/c^2$  i.e very close to ISCO (Inner-most stable circular orbit). The
blackbody component has  temperature in the range of $\sim$ 1.2-1.7
keV. The total flux is the highest during the soft state and hence we
 call it high-soft state.

 A Gaussian iron line component seen in this source is very broad ($FWHM \sim 1.1-1.4$ keV) and strong.  The iron line can also be described by $diskline$ model which includes relativistic effect on the line emission for a non-rotating blackhole \citep{Fabian89}. We fix the emissivity index at $-3$ and the disc inclination at 60$^\circ$. The equivalent width of the iron line is in the range of $\sim$ 100-160 eV. No correlation between iron line flux and total flux is observed. Modeling of the iron line with  $diskline$ model suggests that  the disc is truncated. The truncation radius is anticorrelated with the total flux and varies from 6.8  to 25 $GM/c^2$ as the total flux decreases.

\section{Discussion}

In this work we present broad band spectroscopy (1-80 keV) of the atoll source 4U 1636-536 using combined NuSTAR and Swift-XRT data. We use a spectral model which has emission component from a standard thin accretion disc,a single temperature blackbody from the hot surface of neutron star and Comptonized emission from the corona. An additional relativistically smeared iron line is also present in all the observations.  The model presented in this paper is very similar to the model adopted for this source by \citet{Sanna13}  and \citet{Lyu14}. However, \citet{Sanna13}  use an additional reflection component modeled with BBREFL (reflection of boundary layer photons by accretion disc). 
The source was found in two different states during the observations:
Soft and Hard. Similar state transitions have been reported by Sanna et
al (2013) using 6 XMM plus RXTE observations. During the soft state  
electron temperature was low  and optical depth was high. In the hard
state, electron temperature was  higher and optical depth was lower
compared to the soft state. The spectra in the soft state (Obs-2) and in the hard state (Obs-4) are plotted in Figure 4 to show the state transition.

In the soft spectral state of the source we also detect non-thermal
tail with photon index $\sim$ 2.4. The source flux was highest in this
state (4.36 $\times$ 10$^{-9} ergs/s/cm^2)$. The non-thermal component
contributes 10\% of the unabsorbed 0.1-100 keV flux. Power-law components
with photon indices $\sim$ 2-3 contributing 30-80 \% of the total flux
have been reported during Steep Power-Law state  (SPL) of blackhole
candidates \citep{Remillard06}.  However, a  blackbody emission, probably
coming from the boundary layer around neutron star, with temperature in
the range 1.2-1.7 keV is also present in the spectra of this source. The
presence of boundary layer component suggests that  spectral state seen
in this source is  different from the SPL state of blackhole candidates.
A hard tail has been previously reported in this source using the INTEGRAL
observation \citep{Fiocchi06}. Similar hard tails have been seen previously
in  Z-sources GX 17+2 \citep{disalvo00}, Cyg X-2 \citep{disalvo02},
Sco X-1 \citep{damico1}, GX 349+2 \citep{disalvo01} and atoll sources
4U 1705-44 \citep{Pirano07} and 4U 1728-34 \citep{Tarana11}.

It has been proposed that hybrid thermal/non-thermal population of 
electrons in the corona can produce such a hard tail in the neutron
star and blackhole binaries \citep{Poutanen98}. Alternate mechanism
such as  Comptonization in bulk motion of matter close to the compact
object has also been proposed \citep{Titarchuk98,Ebisawa96}. However,
\citet{Farinelli07} suggested  that at high accretion rate, radiation
pressure due to emission from neutron star surface can slow down  the
bulk motion of matter causing quenching of bulk Comptonization. Hence
the observation of a hard tail at the highest accretion rate suggests that most
probably origin of hard tail due to bulk Comptonization is not favourable.

We have detected a broad iron line during all the four observations. The
iron emission line profile is well described  by a relativistic disc line
assuming a Schwarzschild metric.  Iron line equivalent width does not show
a clear correlation with the flux of Comptonized component (see Table 3) suggesting that
dependence may be complex. The inner disc radius obtained from Fe line
fit is in the range of $6.5-25 GM/c^2$ and is  anti-correlated with the observed
0.1-100 keV flux (see Figure 5) . In the soft state it almost extends ($6.5 GM/c^2)$
upto inner most stable circular orbit. Hence our finding suggests that
inner edge of the disc moves inwards as accretion rate increases.
We also find that as total observed flux increases contribution of disc flux to the total also increases. The increase in disc fraction implies increase in the seed photon supply for inverse Compton scattering. Hence the corona should cool down and  should become compact and optically thick. This is supported by the observed fact that the electron temperature decreases and optical depth increases with increase in the total flux (see Figure 5). 
\section{Conclusion}
Investigation of broad band X-ray spectra of the source revealed the presence of a non-thermal component and a broad iron line. The source exhibited flux and state variations during the observations. The spectral parameters showed significant evolution as the  persistent flux level of source varied. Finally, we have proposed a scenario which explains the spectral evolution of the source.
\section*{Acknowledgments} This research has made use of data and/or software provided by  High Energy Astrophysics Science Archive Research Center (HEASARC).  We thank Dr. Anil Agarwal, GD, SAG, Mr. Subramanya Udupa. DD, CDA and Dr. M. Annadurai, Director, ISAC for encouragement and continuous support to carry out this research.

\end{document}